\documentclass[aps,prc,floatfix,superscriptaddress,twocolumn]{revtex4}

\usepackage{amssymb}
\usepackage{amsmath}
\usepackage{bm}
\usepackage{graphicx}

\begin{document}

\title{Measurement of transparency ratios for protons from short-range correlated pairs}

\newcommand*{\TELAVIV}{Tel Aviv University, Tel Aviv 69978, Israel}
\newcommand*{\TELAVIVindex}{41}
\affiliation{\TELAVIV}
\newcommand*{\UTFSM}{Universidad T\'{e}cnica Federico Santa Mar\'{i}a, Casilla 110-V Valpara\'{i}so, Chile}
\newcommand*{\UTFSMindex}{36}
\affiliation{\UTFSM}
\newcommand*{\ODU}{Old Dominion University, Norfolk, Virginia 23529}
\newcommand*{\ODUindex}{29}
\affiliation{\ODU}
\newcommand*{\MIT}{Massachusetts Institute of Technology, Cambridge, Massachusetts 02139, USA}
\newcommand*{\MITindex}{42}
\affiliation{\MIT}
\newcommand*{\ANL}{Argonne National Laboratory, Argonne, Illinois 60439}
\newcommand*{\ANLindex}{1}
\affiliation{\ANL}
\newcommand*{\ASU}{Arizona State University, Tempe, Arizona 85287-1504}
\newcommand*{\ASUindex}{2}
\affiliation{\ASU}
\newcommand*{\CSUDH}{California State University, Dominguez Hills, Carson, CA 90747}
\newcommand*{\CSUDHindex}{3}
\affiliation{\CSUDH}
\newcommand*{\CANISIUS}{Canisius College, Buffalo, NY}
\newcommand*{\CANISIUSindex}{4}
\affiliation{\CANISIUS}
\newcommand*{\CMU}{Carnegie Mellon University, Pittsburgh, Pennsylvania 15213}
\newcommand*{\CMUindex}{5}
\affiliation{\CMU}
\newcommand*{\CUA}{Catholic University of America, Washington, D.C. 20064}
\newcommand*{\CUAindex}{6}
\affiliation{\CUA}
\newcommand*{\SACLAY}{CEA, Centre de Saclay, Irfu/Service de Physique Nucl\'eaire, 91191 Gif-sur-Yvette, France}
\newcommand*{\SACLAYindex}{7}
\affiliation{\SACLAY}
\newcommand*{\CNU}{Christopher Newport University, Newport News, Virginia 23606}
\newcommand*{\CNUindex}{8}
\affiliation{\CNU}
\newcommand*{\UCONN}{University of Connecticut, Storrs, Connecticut 06269}
\newcommand*{\UCONNindex}{9}
\affiliation{\UCONN}
\newcommand*{\EDINBURGH}{Edinburgh University, Edinburgh EH9 3JZ, United Kingdom}
\newcommand*{\EDINBURGHindex}{10}
\affiliation{\EDINBURGH}
\newcommand*{\FU}{Fairfield University, Fairfield CT 06824}
\newcommand*{\FUindex}{11}
\affiliation{\FU}
\newcommand*{\FIU}{Florida International University, Miami, Florida 33199}
\newcommand*{\FIUindex}{12}
\affiliation{\FIU}
\newcommand*{\FSU}{Florida State University, Tallahassee, Florida 32306}
\newcommand*{\FSUindex}{13}
\affiliation{\FSU}
\newcommand*{\Genova}{Universit$\grave{a}$ di Genova, 16146 Genova, Italy}
\newcommand*{\Genovaindex}{14}
\affiliation{\Genova}
\newcommand*{\GWUI}{The George Washington University, Washington, DC 20052}
\newcommand*{\GWUIindex}{15}
\affiliation{\GWUI}
\newcommand*{\ISU}{Idaho State University, Pocatello, Idaho 83209}
\newcommand*{\ISUindex}{16}
\affiliation{\ISU}
\newcommand*{\INFNFE}{INFN, Sezione di Ferrara, 44100 Ferrara, Italy}
\newcommand*{\INFNFEindex}{17}
\affiliation{\INFNFE}
\newcommand*{\INFNFR}{INFN, Laboratori Nazionali di Frascati, 00044 Frascati, Italy}
\newcommand*{\INFNFRindex}{18}
\affiliation{\INFNFR}
\newcommand*{\INFNGE}{INFN, Sezione di Genova, 16146 Genova, Italy}
\newcommand*{\INFNGEindex}{19}
\affiliation{\INFNGE}
\newcommand*{\INFNRO}{INFN, Sezione di Roma Tor Vergata, 00133 Rome, Italy}
\newcommand*{\INFNROindex}{20}
\affiliation{\INFNRO}
\newcommand*{\ORSAY}{Institut de Physique Nucl\'eaire ORSAY, Orsay, France}
\newcommand*{\ORSAYindex}{21}
\affiliation{\ORSAY}
\newcommand*{\ITEP}{Institute of Theoretical and Experimental Physics, Moscow, 117259, Russia}
\newcommand*{\ITEPindex}{22}
\affiliation{\ITEP}
\newcommand*{\JMU}{James Madison University, Harrisonburg, Virginia 22807}
\newcommand*{\JMUindex}{23}
\affiliation{\JMU}
\newcommand*{\KNU}{Kyungpook National University, Daegu 702-701, Republic of Korea}
\newcommand*{\KNUindex}{24}
\affiliation{\KNU}
\newcommand*{\LPSC}{LPSC, Universite Joseph Fourier, CNRS/IN2P3, INPG, Grenoble, France}
\newcommand*{\LPSCindex}{25}
\affiliation{\LPSC}
\newcommand*{\UNH}{University of New Hampshire, Durham, New Hampshire 03824-3568}
\newcommand*{\UNHindex}{26}
\affiliation{\UNH}
\newcommand*{\NSU}{Norfolk State University, Norfolk, Virginia 23504}
\newcommand*{\NSUindex}{27}
\affiliation{\NSU}
\newcommand*{\OHIOU}{Ohio University, Athens, Ohio  45701}
\newcommand*{\OHIOUindex}{28}
\affiliation{\OHIOU}
\newcommand*{\RPI}{Rensselaer Polytechnic Institute, Troy, New York 12180-3590}
\newcommand*{\RPIindex}{30}
\affiliation{\RPI}
\newcommand*{\URICH}{University of Richmond, Richmond, Virginia 23173}
\newcommand*{\URICHindex}{31}
\affiliation{\URICH}
\newcommand*{\ROMAII}{Universita' di Roma Tor Vergata, 00133 Rome Italy}
\newcommand*{\ROMAIIindex}{32}
\affiliation{\ROMAII}
\newcommand*{\MSU}{Skobeltsyn Nuclear Physics Institute, 119899 Moscow, Russia}
\newcommand*{\MSUindex}{33}
\affiliation{\MSU}
\newcommand*{\MSUa}{Physics Department, Moscow State University, 119899 Moscow, Russia}
\newcommand*{\MSUaindex}{43}
\affiliation{\MSUa}
\newcommand*{\SCAROLINA}{University of South Carolina, Columbia, South Carolina 29208}
\newcommand*{\SCAROLINAindex}{34}
\affiliation{\SCAROLINA}
\newcommand*{\JLAB}{Thomas Jefferson National Accelerator Facility, Newport News, Virginia 23606}
\newcommand*{\JLABindex}{35}
\affiliation{\JLAB}
\newcommand*{\GLASGOW}{University of Glasgow, Glasgow G12 8QQ, United Kingdom}
\newcommand*{\GLASGOWindex}{37}
\affiliation{\GLASGOW}
\newcommand*{\VIRGINIA}{University of Virginia, Charlottesville, Virginia 22901}
\newcommand*{\VIRGINIAindex}{38}
\affiliation{\VIRGINIA}
\newcommand*{\WM}{College of William and Mary, Williamsburg, Virginia 23187-8795}
\newcommand*{\WMindex}{39}
\affiliation{\WM}
\newcommand*{\YEREVAN}{Yerevan Physics Institute, 375036 Yerevan, Armenia}
\newcommand*{\YEREVANindex}{40}
\affiliation{\YEREVAN}

\newcommand*{\NOWINFNGE}{INFN, Sezione di Genova, 16146 Genova, Italy}
\newcommand*{\NOWMSU}{Skobeltsyn Nuclear Physics Institute, 119899 Moscow, Russia}
\newcommand*{\NOWORSAY}{Institut de Physique Nucl\'eaire ORSAY, Orsay, France}
\newcommand*{\NOWROMAII}{Universita' di Roma Tor Vergata, 00133 Rome
  Italy}
\newcommand*{\NOWJLAB}{Thomas Jefferson National Accelerator Facility, Newport News, Virginia 23606}

\author{O.~Hen}
\affiliation{\TELAVIV}
\author {H.~Hakobyan} 
\affiliation{\UTFSM}
\affiliation{\YEREVAN}
\author{R.~Shneor}
\affiliation{\TELAVIV}
\author{E.~Piasetzky}
\affiliation{\TELAVIV}
\author{L.B.~Weinstein}
\affiliation{\ODU}
\author {W.K.~Brooks} 
\affiliation{\UTFSM}
\affiliation{\JLAB}
\author{S. May-Tal~Beck}
\affiliation{\TELAVIV}
\author{S. ~Gilad}
\affiliation{\MIT}
\author{I.~Korover}
\affiliation{\TELAVIV}
\author{A.~Beck}
\affiliation{\TELAVIV}
% ...

\collaboration{The CLAS Collaboration}
\noaffiliation

\date{\today}

\begin{abstract}

  Nuclear transparency, $T_p(A)$, is a measure of the average
  probability for a struck proton to escape the nucleus without
  significant re-interaction. Previously, nuclear transparencies were
  extracted for quasi-elastic $A(e,e'p)$ knockout of protons with
  momentum below the Fermi momentum, where the spectral functions are
  well known. In this paper we extract a novel observable, the
  transparency ratio, $T_p(A)/T_p(^{12}\rm{C})$, for knockout of
  high-missing-momentum protons from the breakup of short range
  correlated pairs ($2N$-SRC) in $\rm{Al}$, $\rm{Fe}$ and $\rm{Pb}$
  nuclei relative to $\rm{C}$. The ratios were measured at momentum
  transfer $Q^2\ge1.5$ (GeV/c)$^2$ and $x_B\ge1.2$ where the reaction
  is expected to be dominated by electron scattering from
  $2N$-SRC. The transparency ratios of the knocked-out protons coming
  from $2N$-SRC breakup are $20 - 30\%$ lower than those of previous
  results for low missing momentum. They agree with Glauber
  calculations and agree with renormalization of the previously
  published transparencies as proposed by recent theoretical
  investigations. The new transparencies scale as $A^{-1/3}$, which is
  consistent with dominance of scattering from nucleons at the nuclear
  surface.
 \end{abstract}

\maketitle


Nuclear transparency, $T(A)$, is defined as the ratio of the cross
section per nucleon for a process on a bound nucleon in the nucleus to
that from a free nucleon. Conventionally, for protons, $T_p(A)$ has
been extracted as the ratio of the measured $A(e,e'p)$ quasi-elastic
(QE) cross section to the calculated Plane-Wave Impulse Approximation
(PWIA) cross section, which does not include Final State Interactions
(FSI). The experimental cross sections are typically integrated over
proton missing momenta below the Fermi momentum ($|P_{miss}|\le
k_F\approx250$ MeV/c), and missing energy, $E_{miss}$, below 80 MeV
corresponding to knockout of mean-field
protons~\cite{ONeill95,Garino92,Abbott98,Garrow02}.  ($\vec P_{miss} =
\vec q - \vec P_p$ and $E_{miss} = \omega - T_p$, where $\vec q$ and
$\omega$ are the momentum and energy transfer of the virtual photon
and $P_p$ and $T_p$ are the momentum and kinetic energy of the
outgoing proton, respectively.)  For a recent review, see
\cite{Dutta12}.

Two-nucleon short-range correlations (2N-SRC) are pairs of nucleons
with high momentum ($\vec{p}_1$, $\vec{p}_2$) that balance each
other. The pair has high relative momentum
($\vec{p}_{rel}=\frac{\vec{p}_1-\vec{p}_2}{2}$) and low center of mass
momentum ($\vec{p}_{c.m.}=\vec{p}_1+\vec{p}_2$), where high and low is
relative to the Fermi momentum. $2N$-SRC consist mainly of
neutron-proton pairs and dominate the tail ($|P|\ge k_F$) of the nuclear
momentum distribution for all
nuclei~\cite{Shneor07,Subedi08,Tang03,Piasetzky06,Day87,egiyan03,egiyan06,fomin12,Frankfurt88,Frankfurt93,Ciofi}.

For the extraction of nuclear transparency from the $A(e,e'p)$ quasi-elastic data, the $2N$-SRC are an obstacle since they remove a fraction of the single-particle strength beyond the missing momentum and energy integration range.  This removed strength is difficult to accurately ascertain and therefore introduces uncertainty to the absolute value of  $T_p(A)$. Published experimental data, following~\cite{ONeill95}, used large correction factors ($1.11\pm0.03$, $1.22\pm0.06$, and $1.28\pm0.10$, for $^{12}\rm{C}$, $^{56}\rm{Fe}$, and $^{197}\rm{Au}$, respectively). These are larger than indicated by more recent calculations~\cite{frankfurt00, Lava04}. This is believed to be the main reason for the discrepancy between the measured  $T_p(A)$ transparencies and calculations using the Glauber approximation to describe the FSI of the outgoing struck proton with the residual nucleus~\cite{frankfurt00, Lava04}.

In this paper we avoid the necessity of using hybrid measured-to-calculated ratios and bypass the uncertainty due to the $2N$-SRC correction factors. We present the transparency ratios of $T_p(A)/T_p(^{12}\rm{C})$, where $A$ stands for $^{27}\rm{Al}$, $^{56}\rm{Fe}$, and $^{208}\rm{Pb}$. These ratios are determined for high missing momentum protons knocked out from the breakup of two-nucleon short-range correlated pairs.


The data presented here were collected as part of the EG2 run period that took place in 2004 in Hall B of the Thomas Jefferson National Accelerator Facility (Jefferson Lab), using a 5.014 GeV unpolarized electron beam and the CEBAF Large Acceptance Spectrometer (CLAS)~\cite{Mecking03}. The analysis was carried out as part of the Jefferson Lab Hall B Data-Mining project~\cite{WeinsteinDOE}. 

CLAS uses a toroidal magnetic field (with electrons bending towards the beam line) and six independent sets of drift chambers, time-of-flight (TOF) scintillation counters, Cherenkov counters (CC), and electro-magnetic calorimeters (EC) for charged particle identification and trajectory reconstruction.  The polar angular acceptance is $8^\circ < \theta < 140^\circ$ and the azimuthal angular acceptance is 50\% at small polar angles, increasing to 80\% at larger polar angles. 

We identified electrons and rejected pions by requiring that negative particles produced more than 2.5 photo-electrons in the Cherenkov counter. Additional electron/pion separation was achieved by demanding a correlation between the energy deposited in the inner and outer parts of the EC divided by the momentum of the particle~\cite{Mecking03}. The total energy deposited by electrons in the calorimeter was closely correlated with the electron momentum over the full momentum range. This indicates the electrons are identified cleanly.  We applied fiducial cuts on the angle and momentum of the electrons to avoid regions with steeply varying acceptance close to the magnetic coils of CLAS. 

Protons in CLAS were identified by requiring that the difference
between the measured time-of-flight of positively charged particles
and that calculated from their measured momentum and the proton mass
be less than two standard deviations. This cut clearly separates  protons from pions/kaons up to $p=2.8$ GeV/c.  Due to statistical limitations, we only show data for protons up to $2.4$ GeV/c. 

The kinetic energy of the incoming electron and emerging electron and proton were corrected event-by-event for coulomb distortions using the Effective Momentum Approximation (EMA)~\cite{Aste:2005wc}. Following~\cite{fomin12} we assume an effective electric  potential equal to $75\%$ of the potential produced by unscreened Z charges at the nucleus center. This amounts to a $3,5,10$ and $20$ MeV correction for $^{12}$C, $^{27}$Al, $^{56}$Fe and $^{208}$Pb, respectively.

\begin{figure}[tbp]
\includegraphics[width=8.5cm, height=6.5cm]{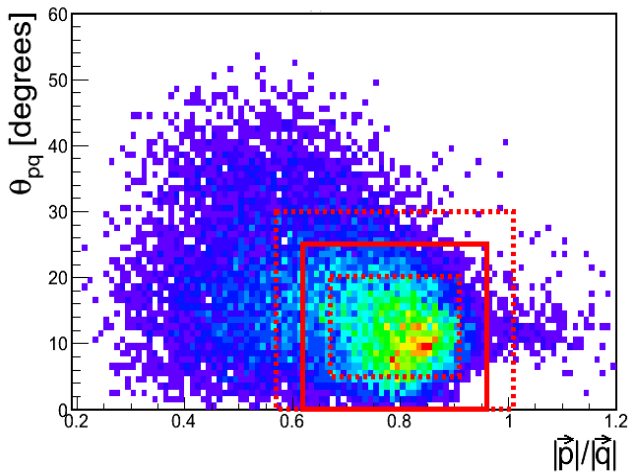}
\caption{The relative angle between the detected proton and the momentum transfer vector $\vec{q}$  versus the ratio of the detected proton momentum and the momentum transfer ($|{\vec{p}}_{p}|/|\vec{q}|$). Only $^{12}{\rm C}(e,e'p)$ events with $x_B\ge1.2$ and $300\le|P_{miss}|\le1000$ MeV/c are shown. The solid / dashed boxes (red online) shows the cuts applied to select leading protons. See table~\ref{tab:cuts} for details.}
\label{fig:LeadingProton}
\end{figure}

	The EG2 run period used a specially designed target setup, consisting of an approximately 2-cm LD$_2$ cryotarget followed by one of six independently-insertable solid targets ranging in thickness  from 0.16 to 0.38 g/cm$^2$ (thin and thick Al, Sn, $\rm{C}$, $\rm{Fe}$, and $\rm{Pb}$, all in natural isotopic abundance)~\cite{Hakobyan08}.  The LD$_2$ target cell and the solid targets were separated by about 4 cm. We selected events with particles scattering from the solid targets by reconstructing the intersections of their trajectories with the beam line. The vertex reconstruction resolution for both electrons and protons was sufficient to unambiguously separate particles originating in the cryotarget and the solid target.  

Cross section ratios for scattering off the solid targets are
defined as the yield ratio, normalised according to the number
of scatterers in the target and the integrated luminosity
accumulated for each target during the experiment. Because all
solid targets were located at the same location along the beam
line and because the $A(e,e'p)$ missing energy and missing momentum
distributions for the different targets were similar, the
detector acceptance effects on the ratios of yields
from different solid targets are negligible in comparison to
our statistical and other systematics uncertainties.

\begin{figure}[tbp]
\includegraphics[width=8.5cm, height=6.5cm]{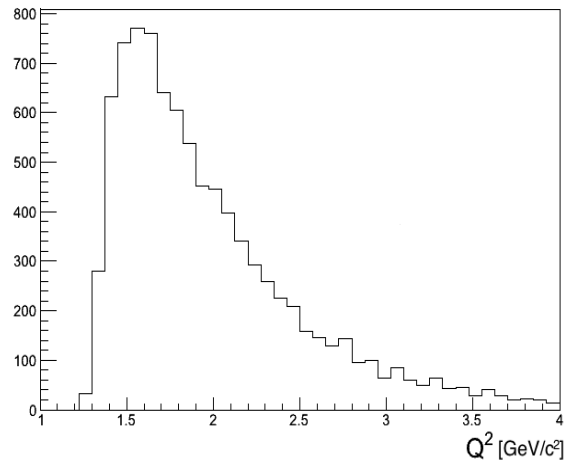}
\caption{$Q^{2}$ distribution for the selected $(e,e'p)$ event sample.}
\label{fig:Q2}
\end{figure}

To identify semi-exclusive $A(e,e'p)$ events dominated by scattering
off $2N$-SRC pairs, one must choose kinematics in which competing
processes are suppressed.  Table~\ref{tab:cuts} lists the cuts
applied and the ranges over which those cuts were varied to determine the
systematic uncertainty.  $Q^2$ and $\omega$
are the four-momentum and energy transfer of the virtual photon,
$x_B=\frac{Q^2}{2m_{N}\omega}$ is the Bjorken scaling variable, and
$m_N$ is the nucleon mass. $\vec P_{miss} = \vec q - \vec P_{p}$ is
the missing momentum which, in the Plane-Wave Impulse Approximation
(PWIA), equals the initial momentum of the proton before it absorbed
the virtual photon. $m_{miss}$ is the reconstructed missing mass for
the $(e,e'p)X$ reaction assuming scattering off a stationary nucleon
pair. $\theta_{pq}$ is the angle between the outgoing proton and the
virtual photon in the lab frame.

\begin{table}[h]
\caption{The $(e,e'p)$ event selection cuts. Also shown is the sensitivity of the transparency ratios to variations in the cuts.}
\raggedright
\begin{tabular}{|c || c |c|c|c|}
\hline
        Cut                                                                 & \multicolumn{4}{|c|}{Cuts Sensitivity}                                      \\ \cline{2-5}
                                                                               & Range                       & Al/C     & Fe/C    & Pb/C      \\
     \hline
     $x_B\ge1.2$                                                    & $\pm0.05$               & $1.4\%$  & $3.2\%$  &  $0.4\%$   \\
     $300\le|\vec{P}_{miss}|\le600$ MeV/c       & $\pm25$ MeV/c*     &  $2.0\%$ & $1.8\%$  &  $2.6\%$   \\
     $\theta_{pq}\le25^\circ$                                & $\pm5^o$**              & $0.6\%$  & $0.3\%$  &  $0.2\%$   \\
     $| |{\vec{P}}_{p}|/|\vec{q}| - 0.79| \le 0.17$ & $\pm0.05$**             &                  &                  &                    \\
     $m_{miss}\le1100$ MeV/c$^2$                  & $\pm50$ MeV/c$^2$   &  $0.5\%$  & $1.1\%$  &  $3.3\%$   \\
    \hline
\end{tabular}
    * The geometric mean of all combinations of $300+25$ MeV/c and $600\pm25$ MeV/c variations are presented.\\
    ** Both leading proton cuts were varied together as shown by the dashed squares in Fig~\ref{fig:LeadingProton}.
\label{tab:cuts}
\end{table}


The cut on $x_B$ is lower than used in inclusive scattering, but the additional cut on $P_{miss}$ ensures the selection of events dominated by scattering off $2N$-SRC pairs, as shown by previous experiments \cite{Shneor07, Subedi08}.  The cut on $m_{miss}$ suppresses the contribution of $\Delta$ excitations and meson production.  The cuts on $|{\vec{P}}_{p}|/|\vec{q}|$ and $\theta_{pq}$ select the struck leading proton (see Fig.~\ref{fig:LeadingProton}).  At most one proton per event passed these cuts, even for events with more than one detected proton.  These cuts combined with the CLAS acceptance result in a momentum transfer distribution that ranges from approximately $1.5$ to $3.5$ (GeV/c)$^2$ (see Fig.~\ref{fig:Q2}).



\begin{figure*}[htpb]
\includegraphics[width=13cm, height=10cm]{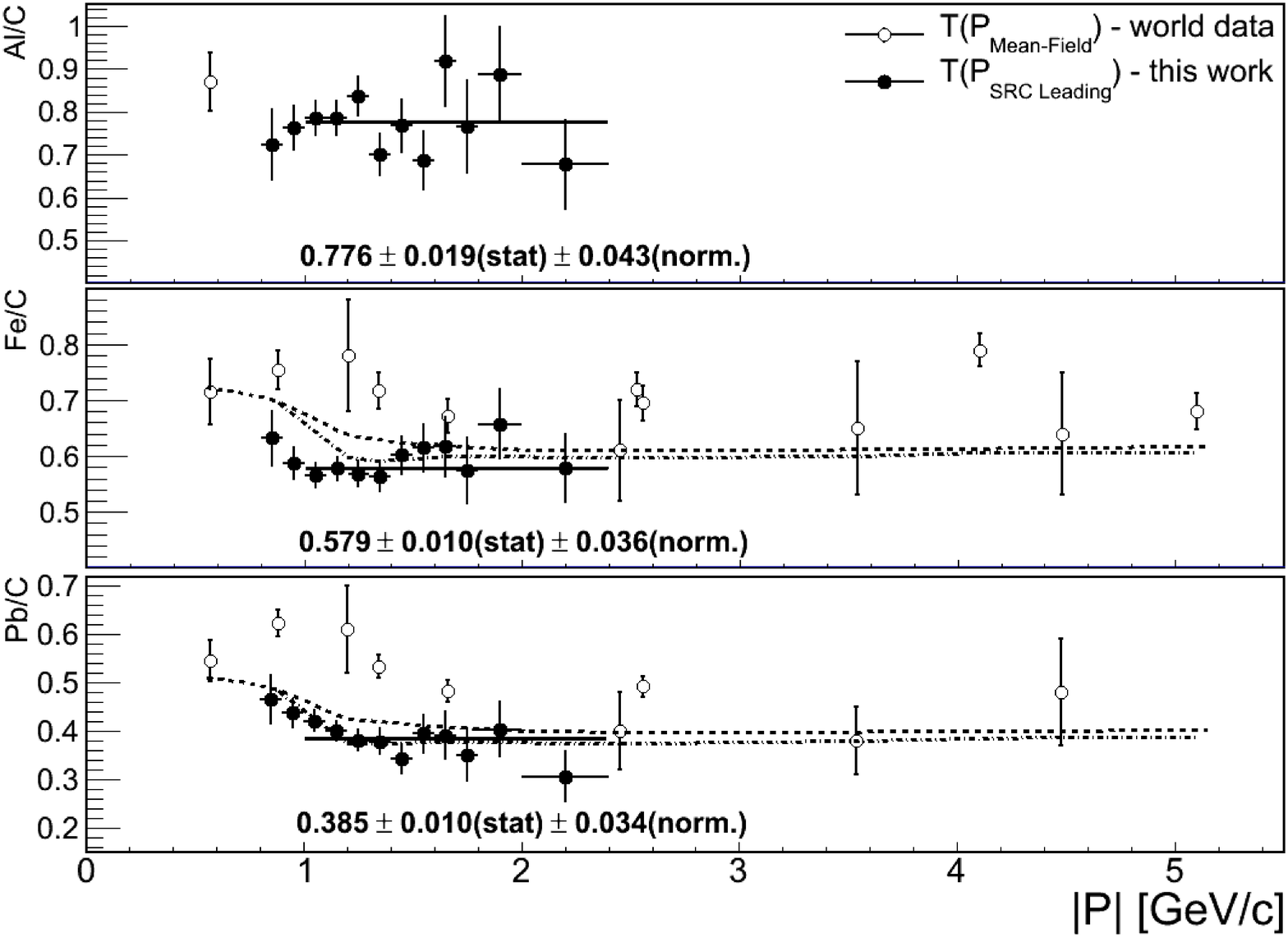}
\caption{The measured transparency ratios for various nuclei with
  respect to carbon of protons from $2N$-SRC pairs (full circles),
  extracted using Eq.~\ref{eq:5a}, shown as a function of the outgoing
  proton momentum. The horizontal error bars represent the integration
  region (bin widths). The solid line is the average transparency and
  the values are shown. The normalization uncertainty is dominated by
  the uncertainties in the SRC scaling factors (see text for
  details). Also shown for comparison are the world data for
  transparency ratios for mean-field proton knockout from
  Ref.~\cite{ONeill95,Garino92,Abbott98,Garrow02} (empty circles),
  extending up to a proton momenta of $~5$ GeV/c. Note that
  Ref.~\cite{Garino92} did not report results for Fe and Pb, we show
  their results for Ni and Ta instead. The results from
  Ref.~\cite{ONeill95,Abbott98} and also~\cite{Garrow02} in the bottom
  panel are for Au rather than Pb. Over the momentum range covered by this
  experiment, the transparency ratios of protons from $2N$-SRC are
  lower than those of mean-field protons by $20-30\%$. Glauber
  calculations are shown as dashed lines~\cite{Pandharipande92} and
  dash-dotted lines~\cite{frankfurt00,Dutta12}. For figure clarity we
  omitted the world data for mean-field transparencies without the SRC
  renormalisation which can be found in~\cite{Dutta12}. }
\label{fig:TransparencyRatios}
\end{figure*}

At these kinematics ($Q^2 > 1.5$ (GeV/c)$^2$,  $x_B >1.2$, and missing momentum  $300\le P_{miss}\le 600$ MeV/c) the nucleon momentum distribution for any given nucleus scales as the number of $2N$-SRC pairs in that nucleus times a common momentum distribution.  This interpretation is strongly supported by both experimental \cite{Shneor07,Subedi08,Tang03,Piasetzky06,Day87,egiyan03,egiyan06,fomin12} and theoretical investigations~\cite{Frankfurt88,Frankfurt93,Ciofi}. The $A(e,e'p)$ cross section in the Plane Wave Impulse Approximation (PWIA) equals a kinematic factor times the elementary electron-proton elastic cross section times the probability of finding a proton at that missing energy and missing momentum.  Under these assumptions, the PWIA cross section ratio for scattering off high-momentum protons from two different nuclei will equal the ratio of the number of $pN$-SRC pairs in the two nuclei (since the other factors all cancel in the ratio).  Since the PWIA cross section does not include the effects of nucleon rescattering as they exit the nucleus, 
we therefore define the proton transparency ratio of any two nuclei in this kinematical regime as the ratio of their measured cross sections scaled by the product of the number of $pN$-SRC pairs:
\begin{equation}
\label{eq:1}
\frac{T_p(A_1)}{T_p(A_2)} = \frac{\sigma_{A_1(e,e'p)}/(N^{A_1}_{np}  +2N^{A_1}_{pp} )}{\sigma_{A_2(e,e'p)}/(N^{A_2}_{np} +2N^{A_2}_{pp}) } ,
\end{equation}
where $\sigma_{A(e,e'p)}$ is the measured quasi-elastic scattering cross section for nucleus $A$, and $N^{A}_{np}$ and $N^{A}_{pp}$ are the average number of $np$ and $pp$ SRC pairs in nucleus $A$.  (The factor of 2 multiplying $N^{A}_{pp}$ reflects the fact that the electron can scatter from either proton in a $pp$ pair.)  

Under minimal assumptions (see Appendix for details), this simplifies to
 \begin{eqnarray}
\label{eq:5a}
\frac{T_p(A_1)}{T_p(A_2)}  =  \frac{1}{a_2(A_1/A_2)} \cdot \frac{\sigma_{A_1(e,e'p)}/A_1}{\sigma_{A_2(e,e'p)}/A_2} ,
\end{eqnarray}
where $a_2(A_1/A_2)$ is the relative number of $2N$-SRC pairs per nucleon in nuclei $A_1$ and $A_2$.  This is exact for isospin symmetric nuclei and should be valid to better than $5\%$ even for asymmetric nuclei such as lead.
The ratios $a_2(A_1/A_2)$ are taken from a compilation of world data on $(e,e')$ cross section ratios at large $Q^2$ and $x_B>1$ including different theoretical corrections~\cite{Hen12}. The values used are: $a_2(^{27}{\rm Al}/^{12}{\rm C})=a_2(^{56}{\rm Fe}/^{12}{\rm C})=1.100\pm0.055$ and $a_2(^{208}{\rm Pb}/^{12}{\rm C})=1.080\pm0.054$. These values are the average of the high precision data of~\cite{fomin12}, with three different sets of theoretical corrections as presented in Table $\rm I$, columns $4-6$ of Ref.~\cite{Hen12}. Notice that the corrections due to the center-of-mass motion of the pair, and their uncertainties, are relevant for the ratios to deuterium and are negligible in the ratio of $A$/$^{12}$C. To be conservative, the uncertainty of $a_2(A_1/A_2)$ was taken to be that of column 6 of Ref.~\cite{Hen12}.


\begin{figure}[t!]
\includegraphics[width=7cm, height=6cm]{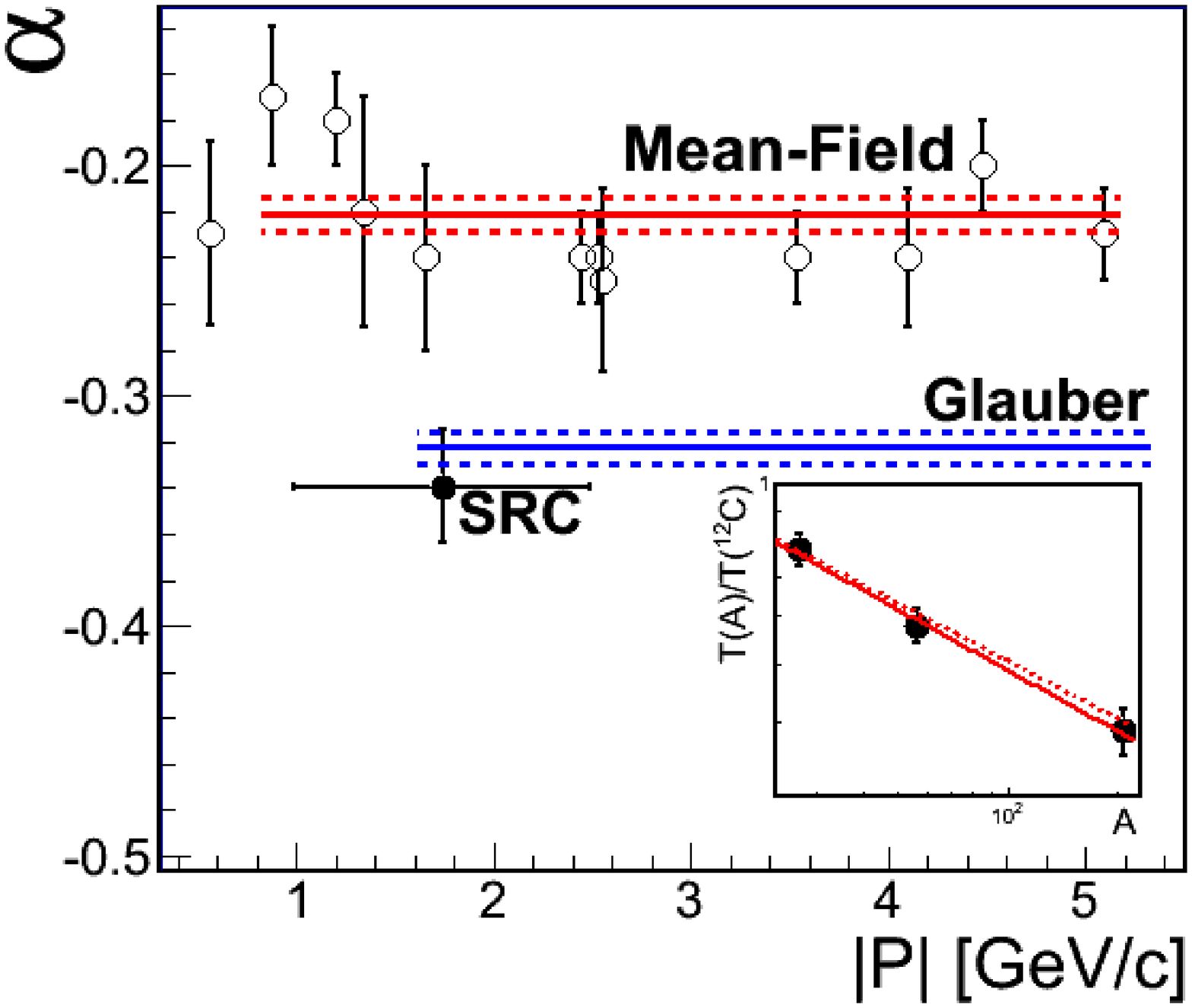}
\caption{The $A$-dependence of the nuclear transparency extracted by fitting the SRC transparency ratios (filled circles) and the mean-field transparency ratios~\cite{Garrow02} (open circles) to $A^{\alpha}$. The solid  line is a constant fit to the world data (red online) and the dashed lines show the $\pm1\sigma$ limits. The dashed-dot line is the Glauber result (blue online) of~\cite{Pandharipande92} .  The insert shows the $A$ dependence of the SRC transparency ratio with the fits to the data  (solid line) and to the Glauber result (dashed line) on a log-log scale. The transparency ratios used are average values shown in Fig~\ref{fig:TransparencyRatios}, where the statistical and normalisation uncertainties were added linearly.  The horizontal error bars indicate the bin width.}
\label{fig:A_Dep}
\end{figure}

The transparency ratios of protons from $2N$-SRC pairs for $^{27}$Al,
$^{56}$Fe, and $^{208}$Pb relative to $^{12}$C, as extracted from the
semi-inclusive $A(e,e'p)$ cross sections in SRC-dominated kinematics
($x_B\ge1.2$, $Q^2\ge 1.5$ (GeV/c)$^2$, and ${P}_{miss}\ge 0.3$
GeV/c), using Eq.~\ref{eq:5a}, are shown in
Fig.~\ref{fig:TransparencyRatios} as a function of the outgoing proton
momentum (which should determine the probability for
re-interacting). The errors shown are statistical only. The $A(e,e'p)$
cross section ratios were corrected for radiative
effects~\cite{Sargsyan_Radiative} in the same way as was done
in~\cite{egiyan03,egiyan06}. The radiative correction to the
transparency ratios was found to equal $\approx7\%$ for all ratios,
with a negligible contribution to the total systematic
uncertainty. The extracted transparency ratios are independent of the
proton momentum for $1.0 \le P_{p} \le 2.4$ GeV/c for each of the
three nuclei.  The average proton transparency, $T_p^{}(A/{\rm C})$,
equals $0.776\pm0.019\pm0.043$ for Al, $0.579\pm0.010\pm0.036$ for Fe
and $0.385\pm0.010\pm0.034$ for Pb. The first uncertainty is
statistical and the second is systematic. The systematic uncertainty
includes the uncertainty in $a_2(A_1/A_2)$ ($5\%$), the sensitivity to
cuts (see table~\ref{tab:cuts}), and the uncertainty of $5\%$ in the
$np$-dominance assumption for the $\rm{^{208}Pb/^{12}C}$ case. The
uncertainty on the integrated luminosity is negligible. The systematic
uncertainty is independent of the proton momentum.  

These transparency
ratios indicate that a high-momentum proton from an SRC pair in iron
is only about 60\% as likely to escape the nucleus as a similar proton
in carbon. This probability ratio for lead is 40\%. These ratios are
20--30\% lower than the corresponding published ratios for mean-field
protons~\cite{ONeill95,Garino92,Abbott98,Garrow02}. Recent theoretical
studies~\cite{frankfurt00, Lava04} claim that the published mean-field
proton transparencies are too high because the PWIA calculations
incorrectly included a correction factor that overestimated the effect
of $2N$-SRC and therefore underestimated the number of available
mean-field protons. This same conclusion was reached from transport
calculations~\cite{Lehr02}. Our measured proton transparency ratios
support this claim.

Following~\cite{ONeill95,Garino92,Abbott98,Garrow02}, the $A$-dependence of the measured transparency ratios was studied by fitting it to $\left({\frac{A}{12}}\right)^{\alpha}$ (see Fig.~\ref{fig:A_Dep}).  Our extracted value of $\alpha=-0.34\pm0.02$ is considerably lower than the average mean-field value of $\alpha=-0.22\pm0.01$~\cite{Garrow02} and is consistent with the Glauber result of $\alpha=-0.322\pm0.007$. 
The observation of $\alpha\approx-1/3$ is consistent with the $T(A)\propto1/r$ attenuation expectation of Ref.~\cite{Lava04}, where $r$ is the nuclear radius, indicating that the reaction is dominated by scattering off the nuclear surface.

In summary, we measured semi-inclusive $A(e,e'p)$ cross section ratios for $^{27}$Al, $^{56}$Fe and $^{208}$Pb nuclei relative to $^{12}$C   at $Q^2\ge1.5$ GeV/c$^{2}$, $x_B\ge1.2$ and $300\le P_{miss}\le600$ MeV/c where knockout of protons from $2N$-SRC should dominate.  We used these cross section ratios to extract the transparency ratios for  protons from the $2N$-SRC breakup. The  proton transparency ratios are independent of  proton momentum and are $20-30\%$ lower than the transparency ratios of mean-field proton knockout.  This difference is consistent with the proposed renormalisation of the mean-field transparencies to properly account for the effects of correlated nucleons~\cite{Lava04,frankfurt00}. See Ref.~\cite{Dutta12} for a comparison of the Glauber calculations to the data, with and without the SRC correction factors.

The $A$-dependence of our measured transparency ratios are steeper than that of mean-field protons~\cite{Garrow02} and consistent with Glauber calculations.  This $A$-dependence is consistent with a simple picture of proton knockout from the nuclear surface, i.e., that protons knocked out from the nuclear volume rescatter.


We acknowledge the efforts of the staff of the Accelerator and Physics
Divisions at Jefferson Lab that made this experiment possible. We are
also grateful for many fruitful discussions with L. L. Frankfurt,
M. Strikman, J. Ryckebusch, W. Cosyn, M. Sargsyan, and C. Ciofi degli
Atti on the formalism and implications of the results. This work was
supported by the U.S. Department of Energy and National Science
Foundation, the Israel Science Foundation, the US-Israeli Bi-National
Science Foundation, the Chilean Comisi\'on Nacional de Investigaci\'on Cient\'ifica y
Tecnol\'ogica (CONICYT) grants FB0821, ACT-119, 1120953, 11121448, and
791100017, the French Centre National de la Recherche Scientifique and
Commissariat a l'Energie Atomique, the French-American Cultural
Exchange (FACE), the Italian Istituto Nazionale di Fisica Nucleare,
the National Research Foundation of Korea, and the United Kingdom's
Science and Technology Facilities Council (STFC). The Jefferson
Science Associates (JSA) operates the Thomas Jefferson National
Accelerator Facility for the United States Department of Energy under
contract DE-AC05-06OR23177.

\section{Appendix}

Inclusive $A(e,e')$ scattering cross section ratios for nuclei $A$ relative to deuterium at $Q^2 > 1.5$ (GeV/c)$^2$ are independent of $x_B$ (scale) for  $1.5 \le x_B\le 1.9$~\cite{Day87,egiyan03,egiyan06,fomin12}. The scaling factor (the value of this cross section ratio), denoted as $a_2(A/d)$, is typically interpreted as a measure of the number of $2N$-SRC pairs per nucleon in nucleus $A$ relative to $d$~\cite{Frankfurt88,Frankfurt93,Ciofi}. When we take the ratio of $a_2(A_1/d)$ and $a_2(A_2/d)$, this gives us $a_2(A_1/A_2)$, the relative number of $2N$-SRC pairs per-nucleon in nuclei $A_1$ and $A_2$.  In this Appendix we will relate this measured quantity to the values $N_{Np}^{A_1}$ used in Eq.~\ref{eq:1}.

In this kinematic region we can assume that the electron scattering cross section from the nucleus is approximately equal to the incoherent sum of electron scattering from the constituent nucleons and therefore is proportional to the number of nucleons times the electron-nucleon cross section. Since at $x_B\ge 1.5$, inclusive electron scattering from nuclei is only sensitive to high-momentum nucleons, this gives: 
\begin{eqnarray}
\label{eq:2}
& & a_2(A_1/A_2)  =   \\
& &  \frac{(N^{A_1}_{np}\cdot(\sigma_{ep}+\sigma_{en})+2N^{A_1}_{pp}\cdot \sigma_{ep}+2N^{A_1}_{nn}\cdot \sigma_{en})/A_1}{(N^{A_2}_{np}\cdot(\sigma_{ep}+\sigma_{en})+2N^{A_2}_{pp}\cdot \sigma_{ep}+2N^{A_2}_{nn}\cdot \sigma_{en})/A_2} ,\nonumber
\end{eqnarray} 
where $\sigma_{eN}$ is the off-shell electron-nucleon elastic scattering cross section and $N^A_{nn}$ is the number of neutron-neutron SRC pairs in nucleus $A$. (For $np$ pairs, the electron can scatter from either the proton or the neutron, so the relevant cross section is $\sigma_{ep}+\sigma_{en}$ and similarly for $nn$ and $pp$ pairs.)

 For isospin symmetric nuclei, we can assume that $N_{nn}=N_{pp}$ and therefore Eq.~\ref{eq:2} simplifies to:
\begin{eqnarray}
\label{eq:3}
 a_2(A_1/A_2) & = & \frac{(\sigma_{ep}+\sigma_{en})\cdot(N^{A_1}_{np}+2N^{A_1}_{pp})/A_1}{(\sigma_{ep}+\sigma_{en})\cdot(N^{A_2}_{np}+2N^{A_2}_{pp})/A_2} \\
& = & \frac{1/A_1}{1/A_2}\cdot\frac{(N^{A_1}_{np}+2N^{A_1}_{pp})}{(N^{A_2}_{np}+2N^{A_2}_{pp})} , \nonumber
\end{eqnarray}
which can be inserted directly in Eq.~\ref{eq:1}.

Even for non-isospin symmetric nuclei Eq.~\ref{eq:3} is reasonably accurate because there are about 20 times more $np$-SRC than $pp$- or $nn$-SRC pairs ($N_{np}\approx 20\times N_{pp}, N_{nn}$)~\cite{Shneor07,Subedi08,Tang03,Piasetzky06}.  If we use the measured value of $N_{np}/N_{pp}=18\pm3$ \cite{Subedi08} and assume that $N^{\rm Pb}_{nn}/N^{\rm Pb}_{pp} = 126^2/82^2 =2.5$, then Eq.~\ref{eq:3} is valid to about 5\%.

Therefore
 we can  rewrite Eq.~\ref{eq:1} as
\begin{eqnarray}
\label{eq:5}
\frac{T_p(A_1)}{T_p(A_2)}  =  \frac{1}{a_2(A_1/A_2)} \cdot \frac{\sigma_{A_1(e,e'p)}/A_1}{\sigma_{A_2(e,e'p)}/A_2} .
\end{eqnarray}
 This is exact for isospin symmetric nuclei and should be valid to better than $5\%$ even for asymmetric nuclei such as lead.

\bibliographystyle{elsarticle-num}

\begin{thebibliography}{23}

\expandafter\ifx\csname natexlab\endcsname\relax\def\natexlab#1{#1}\fi
\expandafter\ifx\csname bibnamefont\endcsname\relax
  \def\bibnamefont#1{#1}\fi
\expandafter\ifx\csname bibfnamefont\endcsname\relax
  \def\bibfnamefont#1{#1}\fi
\expandafter\ifx\csname citenamefont\endcsname\relax
  \def\citenamefont#1{#1}\fi
\expandafter\ifx\csname url\endcsname\relax
  \def\url#1{\texttt{#1}}\fi
\expandafter\ifx\csname urlprefix\endcsname\relax\def\urlprefix{URL }\fi
\providecommand{\bibinfo}[2]{#2}
\providecommand{\eprint}[2][]{\url{#2}}

\bibitem[{\citenamefont{ONeill et~al.}(1995)}]{ONeill95}
\bibinfo{author}{\bibfnamefont{T.~G.}~\bibnamefont{ONeill}} \bibnamefont{et~al.},
  \bibinfo{journal}{Phys. Lett. B} \textbf{\bibinfo{volume}{87}},
  \bibinfo{pages}{351} (\bibinfo{year}{1995}).
  
\bibitem[{\citenamefont{Garino et~al.}(1992)}]{Garino92}
\bibinfo{author}{\bibfnamefont{G.}~\bibnamefont{Garino}} \bibnamefont{et~al.},
  \bibinfo{journal}{Phys. Rev. C} \textbf{\bibinfo{volume}{45}},
  \bibinfo{pages}{780} (\bibinfo{year}{1992}).

\bibitem[{\citenamefont{Abbott et~al.}(2007)}]{Abbott98}
\bibinfo{author}{\bibfnamefont{D.}~\bibnamefont{Abbott}} \bibnamefont{et~al.},
  \bibinfo{journal}{Phys. Rev. Lett} \textbf{\bibinfo{volume}{80}},
  \bibinfo{pages}{5072} (\bibinfo{year}{1998}).
  
\bibitem[{\citenamefont{Garrow et~al.}(1992)}]{Garrow02}
\bibinfo{author}{\bibfnamefont{K.}~\bibnamefont{Garrow}} \bibnamefont{et~al.},
  \bibinfo{journal}{Phys. Rev. C} \textbf{\bibinfo{volume}{66}},
  \bibinfo{pages}{044613} (\bibinfo{year}{2002}).

\bibitem[{\citenamefont{Dutta et~al.}(2012)\citenamefont{Dutta,
  Hafidi and Strikman}}]{Dutta12}
  \bibinfo{author}{\bibfnamefont{D.}~\bibnamefont{Dutta}},
  \bibinfo{author}{\bibfnamefont{K.}~\bibnamefont{Hafidi}}, \bibnamefont{and}
\bibinfo{author}{\bibfnamefont{M.} \bibnamefont{Strikman}},
  \bibinfo{journal}{arXiv:} 
  \bibinfo{pages}{1211.2826} (\bibinfo{year}{2012}).

\bibitem[{\citenamefont{Shneor et~al.}(2007)}]{Shneor07}
\bibinfo{author}{\bibfnamefont{R.}~\bibnamefont{Shneor}} \bibnamefont{et~al.},
  \bibinfo{journal}{Phys. Rev. Lett.} \textbf{\bibinfo{volume}{99}},
  \bibinfo{pages}{072501} (\bibinfo{year}{2007}).

\bibitem[{\citenamefont{Subedi et~al.}(2008)}]{Subedi08}
\bibinfo{author}{\bibfnamefont{R.}~\bibnamefont{Subedi}} \bibnamefont{et~al.},
  \bibinfo{journal}{Science} \textbf{\bibinfo{volume}{320}},
  \bibinfo{pages}{1476} (\bibinfo{year}{2008}).

\bibitem[{\citenamefont{Tang et~al.}(2003)}]{Tang03}
\bibinfo{author}{\bibfnamefont{A.}~\bibnamefont{Tang}} \bibnamefont{et~al.},
  \bibinfo{journal}{Phys. Rev. Lett.} \textbf{\bibinfo{volume}{90}},
  \bibinfo{pages}{042301} (\bibinfo{year}{2003}).

\bibitem[{\citenamefont{Piasetzky et~al.}(2006)}]{Piasetzky06}
\bibinfo{author}{\bibfnamefont{E.}~\bibnamefont{Piasetzky}} \bibnamefont{et~al.},
  \bibinfo{journal}{Phys. Rev. Lett.} \textbf{\bibinfo{volume}{97}},
  \bibinfo{pages}{162504} (\bibinfo{year}{2006}).

\bibitem[{\citenamefont{Day et~al.}(1987)}]{Day87}
\bibinfo{author}{\bibfnamefont{D.}~\bibnamefont{Day}} \bibnamefont{et~al.},
  \bibinfo{journal}{Phys. Rev. Lett.} \textbf{\bibinfo{volume}{59}},
  \bibinfo{pages}{427} (\bibinfo{year}{1987}).

\bibitem[{\citenamefont{Egiyan et~al.}(2003)}]{egiyan03}
\bibinfo{author}{\bibfnamefont{K.}~\bibnamefont{Egiyan}} \bibnamefont{et~al.}
  (\bibinfo{collaboration}{CLAS Collaboration}), \bibinfo{journal}{Phys. Rev.
  C} \textbf{\bibinfo{volume}{68}}, \bibinfo{pages}{014313}
  (\bibinfo{year}{2003}).

\bibitem[{\citenamefont{Egiyan et~al.}(2006)}]{egiyan06}
\bibinfo{author}{\bibfnamefont{K.}~\bibnamefont{Egiyan}} \bibnamefont{et~al.}
  (\bibinfo{collaboration}{CLAS Collaboration}), \bibinfo{journal}{Phys. Rev.
  Lett.} \textbf{\bibinfo{volume}{96}}, \bibinfo{pages}{082501}
  (\bibinfo{year}{2006}).

\bibitem[{\citenamefont{Fomin et~al.}(2012)}]{fomin12}
\bibinfo{author}{\bibfnamefont{N.}~\bibnamefont{Fomin}} \bibnamefont{et~al.},
  \bibinfo{journal}{Phys. Rev. Lett.} \textbf{\bibinfo{volume}{108}},
  \bibinfo{pages}{092502} (\bibinfo{year}{2012}).
  
\bibitem[{\citenamefont{Frankfurt et~al.}(1988)\citenamefont{Frankfurt and
  Strikman}}]{Frankfurt88}
  \bibinfo{author}{\bibfnamefont{L.~L.}~\bibnamefont{Frankfurt}} \bibnamefont{and}
  \bibinfo{author}{\bibfnamefont{M.~I.}~\bibnamefont{Strikman}},
  \bibinfo{journal}{Phys. Rep.} \textbf{\bibinfo{volume}{76}},
  \bibinfo{pages}{215} (\bibinfo{year}{1981}),\bibnamefont{ and}
  \bibinfo{journal}{Phys. Rep.} \textbf{\bibinfo{volume}{160}},
  \bibinfo{pages}{235} (\bibinfo{year}{1988}).

\bibitem[{\citenamefont{Frankfurt et~al.}(1993)\citenamefont{Frankfurt,
  Strikman, Day, and Sargsyan}}]{Frankfurt93}
  \bibinfo{author}{\bibfnamefont{L.~L.}~\bibnamefont{Frankfurt}},
  \bibinfo{author}{\bibfnamefont{M.~I.}~\bibnamefont{Strikman}},
  \bibinfo{author}{\bibfnamefont{D.~B.}~\bibnamefont{Day}}, \bibnamefont{and}
\bibinfo{author}{\bibfnamefont{M.} \bibnamefont{Sargsyan}},
  \bibinfo{journal}{Phys. Rev. C.} \textbf{\bibinfo{volume}{48}},
  \bibinfo{pages}{2451} (\bibinfo{year}{1993}).

  \bibitem[{\citenamefont{Ciofi et~al.}(1994)}]{Ciofi}
  \bibinfo{author}{\bibfnamefont{C.~Ciofi.}~\bibnamefont{degli Atti}}, \bibnamefont{and}
\bibinfo{author}{\bibfnamefont{S.} \bibnamefont{Simula}},
  \bibinfo{journal}{Phys. Lett. B.} \textbf{\bibinfo{volume}{325}},
  \bibinfo{pages}{276} (\bibinfo{year}{1994}),{ and}
  \bibinfo{journal}{Phys. Rev. C.} \textbf{\bibinfo{volume}{53}},
  \bibinfo{pages}{1689} (\bibinfo{year}{1996}).

\bibitem[{\citenamefont{Lava et~al.}(2000)}]{Lava04}
  \bibinfo{author}{\bibfnamefont{P.}~\bibnamefont{Lava}},
  \bibinfo{author}{\bibfnamefont{M.~C.}~\bibnamefont{Martinez}},
  \bibinfo{author}{\bibfnamefont{J.}~\bibnamefont{Ryckebsch}},
\bibinfo{author}{\bibfnamefont{J.~A.} \bibnamefont{Caballero}}, \bibnamefont{and}
  \bibinfo{author}{\bibfnamefont{J.~M.}~\bibnamefont{Udias}},
   \bibinfo{journal}{Phys. Lett. B.} \textbf{\bibinfo{volume}{595}},
  \bibinfo{pages}{177} (\bibinfo{year}{2004}).
  
\bibitem[{\citenamefont{Frankfurt et~al.}(2000)}]{frankfurt00}
  \bibinfo{author}{\bibfnamefont{L.~L.}~\bibnamefont{Frankfurt}},
\bibinfo{author}{\bibfnamefont{M.~I.} \bibnamefont{Strikman}}, \bibnamefont{and}
  \bibinfo{author}{\bibfnamefont{M.}~\bibnamefont{Zhalov}},
   \bibinfo{journal}{Phys. Lett. B.} \textbf{\bibinfo{volume}{503}},
  \bibinfo{pages}{73} (\bibinfo{year}{2001}).

\bibitem[{\citenamefont{Mecking et~al.}(2003)}]{Mecking03}
\bibinfo{author}{\bibfnamefont{B.~A.}~\bibnamefont{Mecking}} \bibnamefont{et~al.},
  \bibinfo{journal}{Nucl. Inst. Meth. A} \textbf{\bibinfo{volume}{503}},
  \bibinfo{pages}{512} (\bibinfo{year}{2003}).
  
\bibitem[{\citenamefont{Weinstein et~al.}(2007)\citenamefont{Weinstein, and Kuhn}}]{WeinsteinDOE}
\bibinfo{author}{\bibfnamefont{L.~B.}~\bibnamefont{Weinstein}},
  \bibnamefont{and}
  \bibinfo{author}{\bibfnamefont{S.~E.}~\bibnamefont{Kuhn}},
  \bibinfo{journal}{``Short Distance Structure of Nuclei: Mining the Wealth of Existing Jefferson Lab Data"}, \bibinfo{volume}{DOE Grant DE-SC0006801.}
  

\bibitem{Aste:2005wc} 
  A.~Aste, C.~von Arx and D.~Trautmann,
  Eur.\ Phys.\ J.\ A {\bf 26}, 167 (2005)
  [nucl-th/0502074].

\bibitem[{\citenamefont{Hakobyan et~al.}(2003)}]{Hakobyan08}
\bibinfo{author}{\bibfnamefont{H.}~\bibnamefont{Hakobyan}} \bibnamefont{et~al.},
  \bibinfo{journal}{Nucl. Inst. Meth. A} \textbf{\bibinfo{volume}{592}},
  \bibinfo{pages}{218-223} (\bibinfo{year}{2008}).

\bibitem[{\citenamefont{Pandharipande et~al.}(2003)}]{Pandharipande92}
\bibinfo{author}{\bibfnamefont{V.~J.}~\bibnamefont{Pandharipande}}, \bibnamefont{and}
  \bibinfo{author}{\bibfnamefont{S.~C.}~\bibnamefont{Pieper}},
  \bibinfo{journal}{Phys. Rev. C} \textbf{\bibinfo{volume}{45}},
  \bibinfo{pages}{791} (\bibinfo{year}{1992}).


\bibitem[{\citenamefont{Hen et~al.}(2012)\citenamefont{Hen,
  Piasetzky, and Weinstein}}]{Hen12}
  \bibinfo{author}{\bibfnamefont{O.}~\bibnamefont{Hen}},
  \bibinfo{author}{\bibfnamefont{E.}~\bibnamefont{Piasetzky}}, \bibnamefont{and}
\bibinfo{author}{\bibfnamefont{L.~B.} \bibnamefont{Weinstein}},
  \bibinfo{journal}{Phys. Rev. C.} \textbf{\bibinfo{volume}{85}},
  \bibinfo{pages}{047301} (\bibinfo{year}{2012}).


  \bibitem[{\citenamefont{Sargsyan_Radiative}(1997)}]{Sargsyan_Radiative}
  \bibinfo{journal}{M. M. Sargsyan, Report No YERPHI-1331-26-91, 1991 (un-published)}
    
\bibitem[{\citenamefont{Lehr et~al.}(2002)}]{Lehr02}
\bibinfo{author}{\bibfnamefont{J.}~\bibnamefont{Lehr}} \bibnamefont{and}
\bibinfo{author}{\bibfnamefont{U.}~\bibnamefont{Mosel}},
  \bibinfo{journal}{Nucl. Phys. A} \textbf{\bibinfo{volume}{699}},
  \bibinfo{pages}{324-327} (\bibinfo{year}{2002}).





  
  
    
\end{thebibliography}

\end{document}